\documentclass[%
reprint,
 amsmath,amssymb,
aps,
]{revtex4-2}

\usepackage{graphicx}

\usepackage{bm} 

\usepackage{dcolumn}

\usepackage{xcolor}
\usepackage[normalem]{ulem}

\newcommand{\edit}[2]{#2}

\begin{document}

\title{Diagnosing Magnetic Fields in Cylindrical Implosions with Oblique Proton Radiography}

\author{P. V. Heuer}
 \email{pheu@lle.rochester.edu}
\affiliation{Laboratory for Laser Energetics, University of Rochester, Rochester, NY 14623, USA}

\author{L. S. Leal}%
\affiliation{Laboratory for Laser Energetics, University of Rochester, Rochester, NY 14623, USA}

\author{J. R. Davies}%
\affiliation{Laboratory for Laser Energetics, University of Rochester, Rochester, NY 14623, USA}

\author{E. C. Hansen}
\affiliation{Laboratory for Laser Energetics, University of Rochester, Rochester, NY 14623, USA}

\author{D. H. Barnak}
\affiliation{Laboratory for Laser Energetics, University of Rochester, Rochester, NY 14623, USA}

\author{J. L. Peebles}
\affiliation{Laboratory for Laser Energetics, University of Rochester, Rochester, NY 14623, USA}

\author{F. Garc\'{i}a-Rubio}
\affiliation{Laboratory for Laser Energetics, University of Rochester, Rochester, NY 14623, USA}

\author{B. Pollock}
\affiliation{Lawrence Livermore National Laboratory, Livermore, CA 94550, USA}

\author{J. Moody}
\affiliation{Lawrence Livermore National Laboratory, Livermore, CA 94550, USA}

\author{A. Birkel}
\affiliation{Plasma Science and Fusion Center, Massachusetts Institute of Technology, Cambridge, MA 02139, USA}

\author{F. H. Seguin}
\affiliation{Plasma Science and Fusion Center, Massachusetts Institute of Technology, Cambridge, MA 02139, USA}

\date{\today}
\begin{abstract}
Two experiments on the OMEGA Laser System used oblique proton radiography to measure magnetic fields in cylindrical implosions with and without an applied axial magnetic field. Although the goal of both experiments was to measure the magnitude of the compressed axial magnetic field in the core of the implosion, this field was obfuscated by two features in the coronal plasma produced by the compression beams: an azimuthal self-generated magnetic field and small length scale, high-amplitude structures attributed to collisionless effects. In order to understand these features, synthetic radiographs are generated using fields produced by 3-D \textit{HYDRA} simulations. These synthetic radiographs reproduce the features of the experimental radiographs with the exception of the small-scale structures. A direct inversion algorithm is successfully applied to a synthetic radiograph, but is only partially able to invert the experimental radiographs in part because some protons are blocked by the field coils. The origins of the radiograph features and their dependence on various experimental parameters are explored. The results of this analysis should inform future measurements of compressed axial magnetic fields in cylindrical implosions.

\end{abstract}

\maketitle



\section{Introduction}

Cylindrical implosions can be used to amplify an applied axial magnetic field via flux compression~\cite{GarciaRubio2018magneticflux}.  If the magnetic flux is completely frozen into the plasma inside the cylinder, the flux compression is proportional to the square of the convergence ratio (CR); however, in practice magnetic diffusion and Nernst advection decrease the field in the compressed core. Cylindrical implosions can be driven by various drivers, including pulsed power Z-pinches~\cite{Slutz2010pulsed} and laser ablation on a cylindrical target~\cite{Davies2017laser}. Flux compression in cylindrical implosions can be used to study fundamental plasma physics in high magnetic fields~\cite{Walsh2022exploring} and is a key feature of the magnetized liner inertial fusion (MagLIF) energy scheme~\cite{Slutz2010pulsed}. 

Previous experiments on the OMEGA Laser System have measured flux compression in cylindrical implosions using proton radiography~\cite{Gotchev2009laser,Knauer2010compressing}. These experiments were followed by the development of the laser-driven MagLIF platform~\cite{Davies2017laser}, which uses smaller-diameter cylindrical targets with a higher maximum convergence and reached maximum convergence more quickly. However, attempts to use proton radiography with this platform to measure the compressed axial magnetic field in the same manner as previous work have so far been unsuccessful, primarily due to the impact on the radiographs of other strong electric and magnetic fields near the target.

\edit{}{The presence of electric and magnetic fields in addition to the compressed axial
magnetic field can interfere with proton radiography measurements. There are often many such fields present in experiments. These additional fields deflect the probing protons, distorting or blocking the signal from the region of interest in the compressed core.}  Electric fields can be created by target charging~\cite{Rygg2008proton,Igumenshchev2014self} or electron pressure gradients, while magnetic fields are primarily generated by the Biermann battery mechanism~\cite{Gao2015precision}.  In cylindrical implosions, radial electron pressure gradients lead to radial electric fields. Fields are also created by the driver: in Z-pinches, the azimuthal pinching field is dominant, while in laser-driven cylindrical implosions, electric and magnetic fields are self-generated in the ablated coronal plasma. Both the azimuthal pinch fields and coronal fields have larger length scales than the compressed core, which particularly impact line-integrated diagnostics like proton radiography.

Two recent experiments at the Omega Laser Facility studied the compression of an applied axial magnetic field in laser-driven cylindrical implosions. The primary goal of both experiments was to use proton radiography~\cite{Kugland2012relation} to measure the magnitude of the compressed axial magnetic field in the core of the implosion. However, this goal was not met because this field was obfuscated by two features in the coronal plasma produced by the compression beams: large-scale azimuthal self-generated magnetic fields and small-scale field structures attributed to collisionless effects in the coronal plasma. 

In this paper we analyze the results of these experiments, determine the origins of the features that obscured the compressed axial field on the radiographs, and explore their dependence on experimental parameters. The design of both experiments is described in Sec.~\ref{sec:setup}, and corresponding 3-D magnetohydrodynamics (MHD) simulations are presented in Sec.~\ref{sec:simulations}. Section~\ref{sec:tracing} describes a particle-tracing algorithm for producing synthetic proton radiographs from the simulated fields. In Sec.~\ref{sec:results}, these synthetic radiographs are used to explain the key features of the experimental radiographs (with the exception of the small-scale structures). In Sec.~\ref{sec:results_inversion}, a direct inversion algorithm is used to retrieve the line-integrated field profile from the synthetic radiographs, but gives limited information on fields in the plasma because some protons are blocked by the field coils. Our conclusions are summarized in Sec.~\ref{sec:conclusion}. 


\section{Experimental Setup\label{sec:setup}}
\begin{figure}
\includegraphics[width = 0.4 \textwidth]{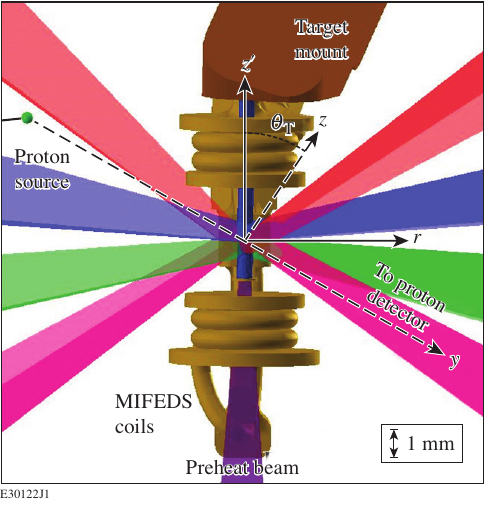}
\caption{\label{setup} A diagram of the setup for Exp. 1, with only a subset of the compression beams shown for clarity. The setup for Exp. 2 is similar.}
\end{figure}



\begin{table}
\begin{tabular}{lccc}

 & Exp. 1  & Exp. 2 & Refs.~\cite{Gotchev2009laser,Knauer2010compressing}           \\ \hline
Cylinder OD ($\mu$m)  & 580  & 640 & 860 \\
\edit{}{Shell thickness ($\mu$m)}  & 20   & 30 & 20 \\
Gas fill (mg/cm$^3$)   & 0.7 (H$_2$)    & N/A  & 0.3 to 0.7 (D$_2$)\\
E$_\text{p}$ (J)           &  180 & N/A  & N/A \\
B$_0$ (T)         &  9    &  5   & 10   \\
$\theta_\text{T}$ (deg)         &  26.5      &  10.8   & 0    \\
Proton timing (ns) & 1.5   & 1.9 & $>$2 \\
Laser pulse length (ns) & 1.5    &   1.5    &1 \\
\edit{}{Convergence ratio} & \edit{}{20}  & \edit{}{20}  & \edit{}{24}  \\
\end{tabular}
\caption{\label{setup_table}Experimental Parmameters for Exp.~1,  Exp.~2, and a previous cylindrical implosion experiment. $E_\text{p}$, $B_0$, and $\theta_\text{T}$ are the preheat beam energy, maximum applied axial magnetic field, and the tilt angle of the proton radiography, respectively.}
\end{table}

Two experiments were conducted (hereinafter Exp.~1 and Exp.~2) utilizing a platform~\cite{Davies2017laser} initially developed for studying laser-driven MagLIF~\cite{Slutz2010pulsed} on the OMEGA Laser System. The setup of Exp.~1 is shown in Fig.~\ref{setup} (Exp.~2 is similar) and the parameters of both experiments are summarized in Table~\ref{setup_table}. In both experiments, the target is a  plastic (CH) cylinder imploded using 40~beams (1.5-ns square-shaped pulse, total energy 16~kJ) with an overlapped intensity of ${\sim}10^{14}$~W/cm$^2$. In Exp.~1 the target is gas filled (14 atm H$_2$), which is preheated by an axial beam (with beam energy E$_\text{p}$) prior to compression as in MagLIF. A set of external coils driven by the MIFEDS (magneto-inertial fusion electrical discharge system)~\cite{Fiksel2015experimental} provides an axial magnetic field. An unmagnetized shot (with the coils in place but not energized) was taken in Exp.~1. Due to the target chamber geometry, in both experiments the proton radiography axis is tilted relative to the target normal by an angle $\theta_\text{T}$. 

In Exp.~2 no preheat beam is used (allowing the setup to be rotated to a lower tilt angle) and the cylinder interior is \edit{}{initially} at vacuum. \edit{}{However, soon after the implosion begins the interior of the cylinder is filled with a CH plasma of unknown density created when the shock driven by the compression beams breaks through the shell (this plasma is visible on x-ray diagnostics)}. A thicker shell is used to delay peak convergence until after the end of the compression pulse. Field compression is possible even without a gas fill because induced currents in the imploding shell are responsible for the compression, not the fill.

\begin{figure}
\includegraphics[width = 0.4\textwidth]{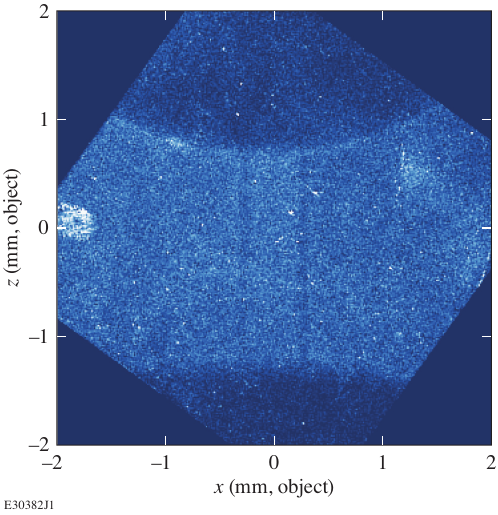}
\caption{\label{undriven_radiograph} \edit{}{A 15 MeV proton radiograph of an undriven and unmagnetized target cylinder shows that protons pass through most of the cylinder without significant attenuation. As a result, only the edges of the cylinder, where the line-integrated density is largest, are visible. The dark circles above and below the cylinder are the shadows of the coils. Bright spots on the image are the result of artifacts on the CR39 detector.}
}
\end{figure}

Proton radiography~\cite{Kugland2012relation} is used to diagnose the fields. A D$^3$He backlighter capsule $l = 11$~mm from the cylinder is imploded by 16~beams to produce ${\sim}3$-MeV and ${\sim}15$-MeV protons (resulting from D-D and D-$^3$He fusion, respectively). Previous experiments with a similar configuration~\cite{Li2006measuring} have characterized the proton source as having a Gaussian radial profile with a full width at half maximum of ${\sim}45$ $\mu$m. The protons pass through the target cylinder walls with negligible scattering \edit{}{(verified by an unmagnetized, uncompressed shot, shown in Fig.~\ref{undriven_radiograph})}, but are deflected by electric and magnetic fields in the vicinity of the target. The protons are then recorded on two CR-39 plates (shielded by $7.5$~$\mu$m of tantalum and separated by $200$~$\mu$m of aluminum  to differentiate between the two proton energies) at a distance $L=270$ mm. \edit{}{In both experiments, the timing of the proton source is chosen to match the peak convergence of the implosion (which is also the peak of neutron production, or \lq bang time\rq) at $t=1.5 \pm 0.1$ ns.} Exp.~2 also included a foil \edit{}{equidistant between the proton source and the cylinder axis}  to block protons, with the intention of isolating protons deflected from the core [the shadow of which is visible in Fig.~\ref{radiographs}(c)].

\edit{}{Proton radiography can also be performed using protons accelerated by the Target Normal Sheath Acceleration (TNSA) mechanism driven by a short pulse laser beam incident on a foil~\cite{Wilks2001energetic}. Previous attempts to probe cylindrical implosions on OMEGA using TNSA protons generated using the OMEGA EP short pulse beam found that the low energy available in this beam in the OMEGA target chamber ($<50$ J) did not produce sufficiently high energy protons for the measurement. Proton radiography is also often performed with a fiducial mesh between the proton source and the object~\cite{Malko2022design, Johnson2022proton}. This technique increases the resolution of the magnetic field measurement, but decreases the spatial resolution (which is then determined by the mesh wire spacing projected to the object plane rather than the detector pixel size). A fiducial mesh was not used in these experiments because the core at peak compression is on the same scale as the finest available meshes. Other techniques for measuring compressed magnetic fields in cylindrical implosions include x-ray dopant~\cite{Walsh2022exploring} and fusion product~\cite{Knapp2015effects} spectroscopy.}

We define two spatial coordinate systems, both with origins at the center of the target cylinder. The first is cylindrical (coordinates $r$, $\phi$, and $z^\prime$) with the $z^\prime$ axis parallel to the cylinder axis. The second is Cartesian (coordinates $x$, $y$, and $z$), with the $y$ axis parallel to the radiography axis and the $x$ axis perpendicular to the cylinder axis.  Times are specified relative to the beginning of the compression laser pulse.

\section{Simulation\label{sec:simulations}}

\begin{figure*}
\includegraphics[width = 1.0 \textwidth]{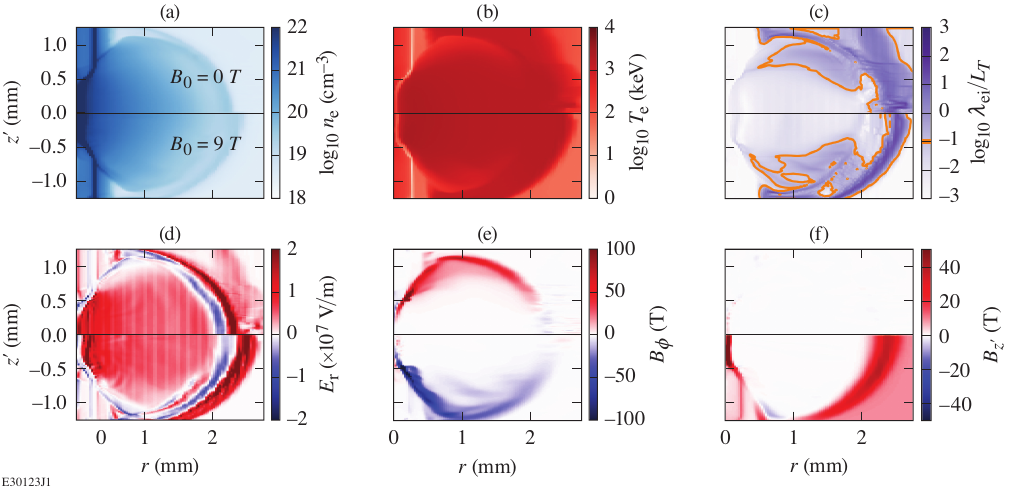}
\caption{\label{hydra} \textit{HYDRA} outputs of the (a) electron density, (b) electron temperature, (c) nonlocality parameter $\lambda_\text{ei}/L_T$, (d) radial electric field, (e) azimuthal magnetic field, and (f) axial magnetic field at $t=1.4$~ns from a simulation of Exp.~1 with an applied axial magnetic field of $B_0 = 9$~T. Simulations of Exp.~2 are qualitatively similar. The top and bottom halves of each plot show the unmagnetized and magnetized simulations respectively. An orange contour in (c) marks the region $\lambda_\text{ei}/L_T > 0.1$, where nonlocal effects are expected to be significant.}
\end{figure*}

The multiphysics radiation-hydrodynamics code \textit{HYDRA}~\cite{Marinak2001three,Farmer2017simulation} is used to perform a series of 3-D simulations of magnetized shots from both experiments, and of the unmagnetized shot from Exp.~1. The experiment is close to axisymmetric but the implosion beam geometry (four rings of ten beams) imposes a weak mode 10 in the implosion. This asymmetry does not significantly affect the results presented here, but 3-D simulations were required to reproduce the measured neutron yields in previous simulations, and are therefore used for these analysis as well. \edit{}{Simulations corresponding to Exp. 2 are initialized with a low density CH plasma inside the cylinder, representing the plasma created when shocks driven by the compression beams break through the shell wall.} Simulations of the unmagnetized shot were run with different flux limiters ($f=0.15$ and $f=0.05$, commonly used values for high and low heat flux, respectively~\cite{Rosen2011role}) to test the extent to which the flux limiter affects the Biermann battery mechanism by modifying the electron temperature gradients~\cite{Ridgers2021inadequacy}. All other simulations were run with a flux limiter of $f=0.15$. 

\textit{HYDRA} includes laser ray-tracing and MHD packages, and here is run using resistive MHD with the addition of some extended MHD features, including the Nernst term. \edit{}{Nernst advection acts to reduce the compressed axial field relative to the frozen-in flux model.} The Righi--Leduc term is included but is reduced in the unmagnetized case (due to numerical instability at low fields). All simulations are initialized on a cylindrical grid (described elsewhere~\cite{Hansen2020neutron}) with identical target and laser-drive configurations corresponding to the experimental configuration. These simulations have previously been demonstrated to model x-ray self-emission, neutron bang time, and neutron-averaged ion temperature from this experimental platform~\cite{Davies2017laser,Hansen2018measuring, Hansen2018optimization,Hansen2020neutron}. In post-processing, the \textit{HYDRA} results are re-gridded (using a nearest-neighbor interpolation) onto a $200 \times 200 \times 200$ Cartesian mesh. The electric field (which is not directly output by \textit{HYDRA}) is calculated as $\bm E = - \nabla P_\text{e}/e n_\text{e}$. Other electric-field sources, such as the Hall term, are found to be negligible.

The simulation results (Fig.~\ref{hydra}) show a coronal plasma expanding away from the axis as the cylinder implodes. Steep density and temperature gradients are present within the ablated plasma and the cylinder. The dominant electromagnetic-field components are a radial electric field $E_\text{r} \sim 10^7$~V/m, an azimuthal magnetic field $B_\phi \sim 50$~T, \edit{}{and in simulations with an applied magnetic field, the compressed axial field $B_z \sim 200$ T}. The orientation of \edit{}{the azimuthal} magnetic field is consistent with generation by the Biermann battery mechanism due to nonparallel density and temperature gradients in the coronal plasma. Runs with a lower flux limiter ($f=0.05$) produce a marginally larger $\nabla T_\text{e}$, but the effect on \edit{}{the fields is small}.

\edit{}{Due to numerical instabilities in the coronal plasma, the simulations are terminated at $t=1.4$~ns, ${\sim} 0.1$~ns before peak convergence. This choice was made to prioritize accurate simulation of the self-generated fields in the coronal plasma, which dominates the experimental proton radiographs. As a result, the final convergence ratio in the simulation ($\text{CR}\sim5$, $\text{ID}\sim 110$~$\mu$m) is much lower than the maximum convergence expected in the experiments ($\text{CR}\sim20$, $\text{ID}\sim 27$~$\mu$m)~\cite{Davies2019inferring, Leal2022effect}. The compressed axial field at the end of the simulation ($B_z \sim 200$~T) is correspondingly smaller than expected in the experiment at peak convergence ($B_z \sim 3$~kT) when the experimental proton radiographs are recorded. The effect of this discrepency on the proton radiographs of the compressed core is discussed in detail in Section~\ref{sec:results_features}. The self-generated fields in the coronal plasma are effectively unchanged over $0.1$~ns, and are therefore still directly comparable between the simulations and the experiment. }

To determine the regions in which nonlocal effects may be significant, the nonlocality parameter $\lambda_\text{ei}/L_T$ is calculated from the \textit{HYDRA} results, where $L_T=T_\text{e}/\nabla T_\text{e}$ is the gradient length scale of the electron temperature $T_e$ and $\lambda_{ei}$ is the electron--ion mean free path $16 \pi \epsilon_0^2 T_\text{e}^2/Z n_\text{e} e^4 \log \Lambda$, where $\epsilon_0$ is the permittivity of free space, $n_\text{e}$ is the electron density, and $e$ is the fundamental charge. \edit{}{The mean ion charge state is $Z=3.5$} and the Coulomb logarithm is $\log \Lambda=8$. Significant nonlocal effects are expected when $\lambda_\text{ei}/L_T > 0.1$~\cite{Bell1981electron}, which is when collisionless effects become important. The results, shown in Fig.~\ref{hydra}(c), show that much of the coronal plasma is either in or approaching the collisionless regime where the fluid approximation is not valid. 

Several mechanisms that could modify the magnetic-field profile were found to be negligible. Nernst advection of self-generated fields could modify the magnetic-field profile (reducing the maximum field in the compressed core)~\cite{Lancia2014topology,Gao2015precision}, but varying the Nernst multiplier from zero to two in our simulations does not significantly change synthetic radiographs generated from the simulated fields (Sec.~\ref{sec:results}). This is because the transport of magnetic flux in the corona is dominated by fluid advection. Similarly, energy transport in the corona is dominated by convection rather than conduction. This is illustrated by the fact that varying the flux limiter in our simulations also does not change the synthetic radiographs. Finally, anomalous resistive dissipation could reduce the magnetic-field strength, but a ${\gtrsim}10^3$ increase in effective electron collision frequency would be required to give significant resistive dissipation on the experimental time scale, which would have further physical consequences (such as increased laser absorption in the corona and reduced thermal conduction) that are not consistent with the experimental results.

\section{Synthetic Charged-Particle Radiography\label{sec:tracing}}

To directly compare simulations to experimental results, synthetic proton radiographs are generated using an open-source particle-tracing algorithm that was developed for the PlasmaPy project as part of this work~\cite{PlasmaPyv07}. The particle-tracing algorithm is initialized with the fields from the \textit{HYDRA} simulations (the ``grid") and parameters specifying a population of protons. A flat-top Gaussian mask is applied to the simulated fields to eliminate edge effects. For the synthetic radiographs presented here, a monoenergetic population of $10^6$ protons is initialized with a uniform velocity distribution in solid angle at the location of the backlighter capsule. The location of the backlighter capsule and orientation of the initial proton velocities can be freely specified relative to the simulated field grid to reproduce the different tilt angles $\theta_\text{T}$. 

The protons are first advanced in a single step from the backlighter capsule to the edge of the grid. Protons whose trajectories never intersect the grid are advanced immediately to the detector plane. The protons that do intersect the grid are advanced through the simulated fields using the energy-conserving Boris push algorithm~\cite{Birdsall2004plasma} with an adaptive time step determined by the local grid-crossing time and gyroperiod experienced by each proton. At each time step, the volume-weighted electric and magnetic fields experienced by each proton are interpolated from the simulated fields at the eight grid points surrounding each particle position. Once all protons have passed through the grid, any protons deflected by ${>}180^\circ$ (such that they are traveling away from the detector plane) are removed. The remaining protons are then advanced in a single step to the detector plane. A synthetic radiograph is created as a 2-D histogram of the final proton positions in the detector plane.

\section{Experimental Results\label{sec:results}}

\begin{figure*}
\includegraphics[width = \textwidth]{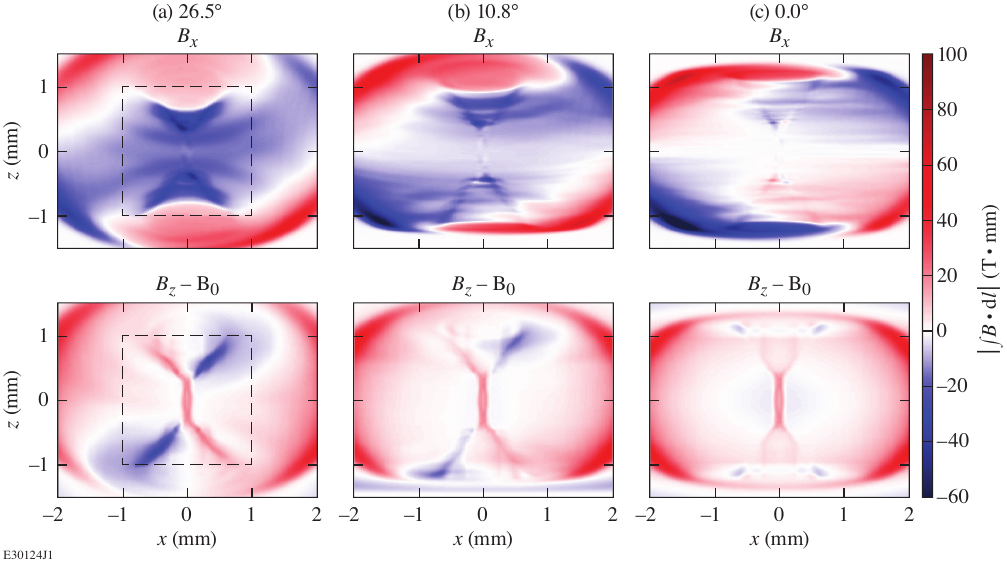}
\caption{\label{line_integrated_field} Horizontal ($B_x$, top) and vertical ($B_z$, bottom) magnetic fields line-integrated along the radiography axis from a simulation of a magnetized shot in Exp.~1 (the fields predicted for Exp. 2 are qualitatively similar). The applied field ($B_0 = 9$ T) is subtracted from $B_z$ for display purposes. The features observed in the experiments are created by the fields in the central region [dashed box on (a)], while the features created by fields at large radii are largely blocked by the coils. \edit{}{At larger $\theta_T$ both} the horizontal and vertical components include a strong contribution from the self-generated azimuthal fields at large $\theta_\text{T}$ \edit{}{[including the 'branches' in $B_z$ in both (a) and (b)]}, but these cancel out at smaller $\theta_\text{T}$ leaving the center to be empty in $B_x$ and dominated by the compressed axial field in $B_z$.} 
\end{figure*}

As protons from the backlighter pass through the experiment they are deflected by an angle that depends on the line-integrated electric and magnetic fields perpendicular to the radiography axis. In the limit of small-angle deflections,
the angular deflection $\alpha  = (\alpha_x \hat x + \alpha_z \hat z)$ of a proton (mass $m_\text{p}$, charge $e$, \edit{}{kinetic} energy $W$) passing through the titled azimuthally symmetric field is~\cite{Kugland2012relation}
\begin{equation}
\alpha = \int  \bigg [ \frac{e \vec E (r, z^\prime) }{2W}  + \frac{e \vec B(r, z^\prime) \times \hat y}{\sqrt{2 m_\text{p} W}} \bigg ] \cdot (\hat x + \hat z) \text{d}y.
\label{deflection_angle}
\end{equation}

Both simulations and end-on x-ray imaging show that the experiment is approximately symmetric in $\phi$. The cylindrical field components in the tilted proton radiography coordinate system are
\begin{align*}
\begin{bmatrix}
B_x\\
B_y\\
B_z
\end{bmatrix} 
&= 
\begin{bmatrix}
B_r \cos \phi - B_\phi \sin \phi \\
B_r \sin \phi \cos \theta_\text{T} + B_\phi \cos \phi \cos \theta_\text{T} - B_{z^\prime} \sin \theta_\text{T} \\
B_r \sin \phi \sin \theta_\text{T} + B_\phi \cos \phi \sin \theta_\text{T} + B_{z^\prime} \cos \theta_\text{T}
\end{bmatrix}. 
\end{align*}
\edit{}{
\begin{align*}
\begin{bmatrix}
E_x\\
E_y\\
E_z
\end{bmatrix} 
&= 
\begin{bmatrix}
E_r \cos \phi - E_\phi \sin \phi \\
E_r \sin \phi \cos \theta_\text{T} + E_\phi \cos \phi \cos \theta_\text{T} - E_{z^\prime} \sin \theta_\text{T} \\
E_r \sin \phi \sin \theta_\text{T} + E_\phi \cos \phi \sin \theta_\text{T} + E_{z^\prime} \cos \theta_\text{T}
\end{bmatrix}. 
\end{align*}
For the configuration described here, where only $B_\phi$, $B_z$, and $E_r$ are non-negligible, the total deflection angles are
\begin{multline}
\alpha_x =   \frac{e}{2W} \int E_r \cos \phi \text{d}y  \\ + \frac{e}{\sqrt{2 m_p W}} \int ( B_\phi \cos \phi \sin \theta_\text{T}  + B_{z^\prime} \cos \theta_\text{T}) \text{d}y
\label{alpha_x}
\end{multline}
\begin{equation}
\alpha_z = - \frac{e}{\sqrt{2 m_p W}} \int B_\phi \sin \phi \text{d}y 
\label{alpha_z}
\end{equation}
}

The line integrals through the simulated fields along the radiography axis for several different values of $\theta_\text{T}$ are shown in Fig.~\ref{line_integrated_field}. When the cylinder tilt angle $\theta_\text{T} = 0^\circ$, 
\edit{}{the $B_\phi$ term in Eq.~\ref{alpha_x} is zero and the $B_\phi$ term in Eq.~\ref{alpha_z} integrates to zero. As a result, to first order the azimuthal magnetic field causes no deflection [Fig.~\ref{line_integrated_field}(c)].} When $\theta_\text{T} > 0^\circ$, an axially uniform cylindrically symmetric $B_\phi$ produces a horizontal ($\hat x$) deflection \edit{}{but no vertical ($\hat z$) deflection.} Since the path length of the line integral also increases proportional to $1/\cos \theta_\text{T}$, $\int B_\phi \text{d}l \propto \tan\theta_\text{T}$. At large tilt angles [Fig.~\ref{line_integrated_field}(a), Exp.~1] strong line-integrated fields near the origin and weaker fields above and below the core are caused by self-generated azimuthal magnetic fields at small and large radius, respectively. For moderate tilts [Fig.~\ref{line_integrated_field}(b), Exp.~2] the fields at small radius largely cancel out but the fields at large radius still contribute to fields above and below the core. \edit{}{When $\theta_\text{T} > 0^\circ$,} axial nonuniformities can result in an additional vertical ($\hat z$) deflections.  This effect could be used to make axially resolved measurement of the azimuthal magnetic field in other experiments, e.g., to measure azimuthal fields in laser--foil interactions where axial proton radiography suffers from scattering in the target~\cite{Campbell2020magnetic}. \edit{}{In some cases with known field geometries, oblique proton radiography can also be used to differentiate between electric and magnetic fields~\cite{Hua2019self}.}

\begin{figure*}
\includegraphics[width = 1.0 \textwidth]{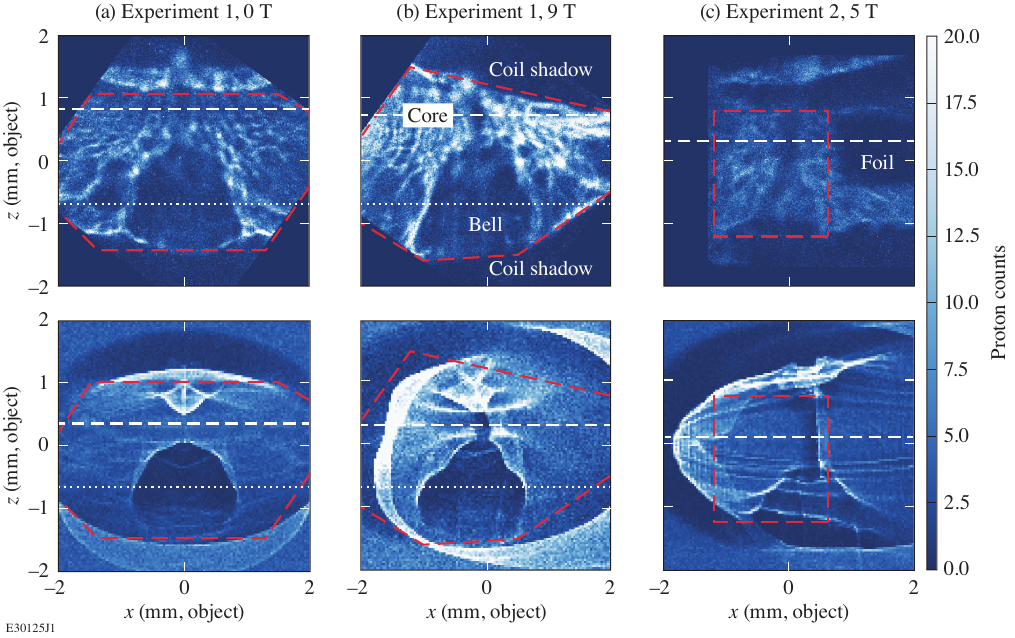}
\caption{\label{radiographs} Experimental radiographs (top) and synthetic radiographs from corresponding simulations (bottom) for three shots with different applied magnetic fields and radiography orientations ($\theta_\text{T}$). \edit{}{Experimental and synthetic radiographs are recorded at and ${\sim}0.1$ ns before peak compression respectively.} The location of lineouts through the core and bell regions are marked with white dashed and dotted lines respectively (these lineouts are plotted in Fig.~\ref{radiograph_lineouts}). \edit{}{The regions outside the red dashed lines on the experimental radiographs are either blocked by the coil features or are off the edges of the detector. The same regions are shown on the synthetic radiographs for comparison.} The synthetic radiographs are scaled to the number of counts in the corresponding experimental radiograph by the ratio $I_{0,\text{synthetic}}/I_{0,\text{experiment}}$.
}
\end{figure*}

\subsection{Radiograph Features\label{sec:results_features}}

In both experiments, radiographs are recorded with both $3$-MeV and $15$-MeV protons. However, the line-integrated fields in the region of interest are sufficiently large to completely block the $3$-MeV protons, so we will restrict our analysis to the $15$-MeV radiographs. These radiographs (Fig.~\ref{radiographs}) contain four primary features, which we will refer to as the coil features, the small-scale structure, the bell feature, and the core feature.

The coil features, regions of low proton flux at the top and bottom of the radiographs, are shadows of the coils that generate the axial magnetic field. The coils block protons whose paths directly intersect them. Some protons are deflected near the edges of the coils (even on the unmagnetized shot), consistent with an electrostatic field at the coil surface caused by an accumulated charge on the coils from ejected electrons. In the magnetized shots, some protons passing near the coils are also deflected by the fringe fields, increasing the size and changing the shape of the coil feature. The curvature of the applied magnetic field also distorts the proton flux on the magnetized shots, leading to the horizontal asymmetry of the coil features. The coils are not included in the particle-tracing algorithm, and consequently the coil shadows do not appear in the synthetic radiographs. Many of the regions of high proton flux far from the center of the plane in the synthetic radiographs are blocked by the coil shadow in the experiments.


The space between the larger image features is filled with small length scale (${\sim}50$ $\mu$m in the object plane) large amplitude fluctuations in the proton flux. These fluctuations correspond to small-scale steep field gradients in the coronal plasma surrounding the implosion. Both the electrons and ions in this region are approximately collisionless [Fig.~\ref{hydra}(c)], so it is possible that these features are caused by kinetic effects such as instabilities or charge-separation fronts. Another possibility is that the coil supports are accumulating charge and these features are created by electrons streaming from imperfections in the coil support surface. These structures are still present in radiographs from Exp.~2, ${\sim}400$ ps after the end of the laser drive. In practice, these peaks represent noise that makes it difficult to pick out the peak corresponding to protons deflected by the compressed axial magnetic field.

Similar small-scale structures have been observed in proton radiographs of spherical implosions~\cite{Rygg2008proton, Zylstra2012using, Seguin2012time} both during and after the laser drive. However, this structure is notably absent in previous experiments that measured compressed fields in magnetized cylindrical implosions~\cite{Gotchev2009laser, Knauer2010compressing}. We conjecture that either the later probe time in these earlier experiments or the relative positions and size of the coils may explain this difference.

The bell feature is a bell-shaped region of depleted proton flux in the bottom half of each radiograph (Fig.~\ref{radiographs}). In Exp.~2 [Fig.~\ref{radiographs}(c)], the bell feature is mostly obscured by the coil shadow. Based on the size of the bell feature, protons from the center must be deflected out by a minimum of $100$~mrad. This deflection is in the nonlinear regime~\cite{Kugland2012relation, Bott2017proton}, which means that a linear inversion to recover the integrated field from the radiograph is not possible and Eq.~(\ref{deflection_angle}) is only a first-order approximation. 

However, as an order-of-magnitude estimate and assuming a length scale of ${\sim}1$~mm, we note that Eq.~(\ref{deflection_angle}) requires either an electric field of $E \sim 10^9$~V/m or a magnetic field of $B \sim 50 $~T to reproduce the observed deflection. Comparing these values to those predicted by the MHD simulation ($E \sim 10^7$~V/m, $B \sim 50$~T) indicates that the magnetic field is responsible. Furthermore, the presence of this feature on the initially unmagnetized implosion (where $B_z \sim 0$~T) indicates that this feature is created by the self-generated azimuthal magnetic field. 

\begin{figure}
\includegraphics[width = 0.45 \textwidth]{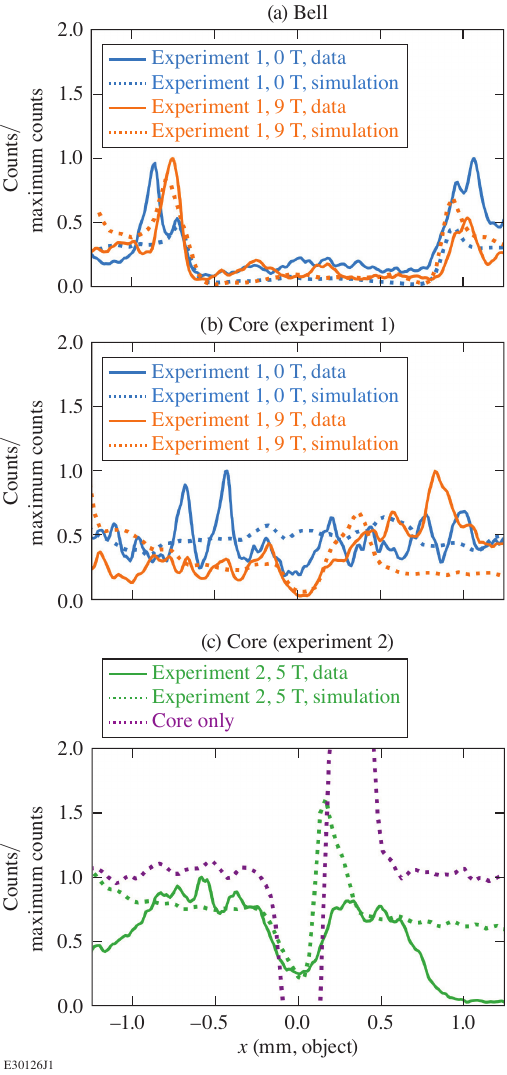}
\caption{\label{radiograph_lineouts} Lineouts through the (a) bell and (b,c) core features from both the experimental (solid) and synthetic (dotted) radiographs from both experiments. Each pair of experimental and synthetic radiographs is normalized to the maximum of the experimental radiograph. The \edit{}{purple} dotted line in (c) is a lineout through an additional synthetic radiograph including only the compressed axial field in the core, and is normalized to its maximum. Since each different applied field shifts the radiograph horizontally by a different amount, each lineout is shifted relative to the others to align the features of interest.
}
\end{figure}

The core feature is a linear region ${\sim}100$~$\mu$m wide of depleted flux above the bell feature that is prominent in both magnetized radiographs but not in the unmagnetized radiograph. The core feature is mostly obscured by the azimuthal magnetic field feature in Exp.~1, but is readily apparent in Exp.~2 because of the smaller cylinder tilt angle. Protons deflected out of the core form an intensity peak to one side of the core feature, which is particularly visible to the right of the depleted region in the synthetic radiographs corresponding to Exp.~2 [Figs.~\ref{radiographs}(c) and~\ref{radiograph_lineouts}(c)]. 

Based on the width (${\sim}100$~$\mu$m) and shape of the core feature, we infer that it is created by fields in the compressed core of the cylindrical implosion and that protons from the center must be deflected out by a minimum of $10$~mrad. By the same order-of-magnitude estimate applied above (now assuming a length scale within the compressed cylinder of ${\sim}25$ $\mu$m), this deflection again requires a minimum of either an electric field of $E {\sim} 10^{10}$~V/m or a magnetic field of $B {\sim} 200$~T. However, the absence of this feature on the initially unmagnetized shot suggests that the compressed axial magnetic field is primarily responsible. The compressed axial field in the simulations ($B_z \sim 200$ T) corresponds to a deflection of ${\sim}400$ $\mu$m in the object plane, which creates a clearly visible peak in the magnetized synthetic radiographs. \edit{}{In the experimental radiographs, which are recorded near peak compression, the field is expected to be ${\sim}$3 kT. This field would produce a peak deflected by ${\sim}2$ mm, which still falls on the detector. However this peak (if it is present) is lost among the many peaks of similar scale in the small-scale structure and therefore cannot be used to make a measurement of the compressed axial magnetic field from the experimental radiograph.}

\edit{}{Near peak compression, the diameter of the compressed core is comparable with the ${\sim}45$ $\mu$m size of the proton source, and as a result the core feature and the peak of deflected protons in the experimental radiograph will be blurred. Even though all protons passing through the core are significantly deflected, convolution of the radiograph with the source profile means that proton flux in the core feature is not zero. Consequently, this proton fluence around the core feature cannot be used to determine the compressed magnetic field. However, the compressed field strength can be measured using the distance of the of the deflected proton peak from the cylinder axis provided this distance is large compared to the source size (which is the case in these experiments). Source blurring is not observed in the synthetic radiographs because, at the time the simulation is terminated, the core is still significantly larger than the proton source.}

\subsection{Quantitative Comparison to Synthetic Radiographs\label{sec:results_quantitative}}

Synthetic radiographs created using the particle-tracing routine (Sec.~\ref{sec:tracing}) include the nonlinear large deflection angle effects neglected in Eq.~(\ref{deflection_angle}). Direct comparisons show that the synthetic radiographs (Fig.~\ref{radiographs}, bottom row) reproduce the bell and core feature observed in the experimental radiographs (Fig.~\ref{radiographs}, top row). The coil shadows are not included in the synthetic radiographs, and the fields that create the small-scale structure are not reproduced by the MHD simulations. To make a quantitative comparison, horizontal lineouts are taken through the center of the bell and core regions. The intensity along the lineout in the image plane, $I$, is approximately related to the gradients of the line-integrated field (in the small-deflection limit)~\cite{Kugland2012relation}: 
\begin{equation}
\frac{I}{I_0} \propto \frac{1}{ |1  + \frac{\partial}{\partial x} \int \vec B \times \vec{\text{d}y}|},
\end{equation}
where $I_0$ is the unperturbed intensity in the image plane (often called the source profile) and the derivative is in the object plane. The ratio $I/I_0$ can therefore be used as a proxy for the line-integrated magnetic-field strength.

The source profile $I_0$ is known \textit{a priori} for the synthetic radiographs. However, estimating this value for the experimental radiographs can be challenging. If the proton source profile is uniform and the radiograph includes a region in which no protons have been deflected in or out, the source profile could be directly measured. If such a region is not available but all deflected protons and depleted regions are included on the radiograph, $I_0$ can be experimentally estimated as the mean proton flux over the entire radiograph. Alternatively, if the proton source is isotropic, $I_0$ could also be estimated as the mean of the proton flux on a second detector positioned to capture undeflected protons.

Unfortunately, in these experiments none of these options are available. The radiographs include no regions without deflections, and some protons are not collected on the detector (either being blocked by the coils or deflected off of the detector). The proton source is isotropic, but a measurement of $I_0$ with a second detector was not possible (also because of the position of the coils). Instead, the experimental $I_0$ is estimated as the mean proton flux in the region where protons are not blocked by the coils. This could be either an overestimate or underestimate, depending on whether the number of protons deflected out of this region exceeds the number deflected in. Tests with synthetic radiographs indicate that the error in this estimate could be as high as 50\%. Therefore, while this normalization is used to scale the experimental data to the synthetic radiographs in Figs.~\ref{radiographs} and \ref{radiograph_lineouts} for display purposes, we will focus our quantitative comparison only on the location (rather than the magnitude) of the features.

Lineouts through the bell feature [Fig.~\ref{radiograph_lineouts}(a)] show that its width does not change significantly between the magnetized and unmagnetized shots, and that in both cases the width is accurately reproduced by the synthetic radiographs. This supports our conclusion that this feature is produced by the self-generated azimuthal magnetic fields, which in the simulations are mostly unchanged by the addition of the applied field. The difference in the depth of the bell feature between the synthetic  and experimental radiographs is within the error in the experimental $I/I_0$, which is dominated by the uncertainty in $I_0$. 

Lineouts through the core feature [Fig.~\ref{radiograph_lineouts}(b,c)] on the magnetized synthetic radiographs show a clear dip to the left of a corresponding peak, corresponding to deflections by the compressed axial magnetic field. As expected, no dip is visible on the unmagnetized synthetic radiograph.  The effect of the compressed axial magnetic field is illustrated by a lineout through a fourth synthetic radiograph [Fig.~\ref{radiograph_lineouts}(c), dotted line, purple] which includes only the axial field in the core region but still reproduces this feature. In the full simulated fields the protons deflected out of this core region are further scattered by other field components, reducing the prominence of the resulting peak.

The experimental radiographs all contain many maxima and minima, most of which belong to the small-scale structure. The two magnetized experimental radiographs both show a more prominent dip at $x=0$. All three experimental radiographs show many peaks (mostly belonging to the small structure), but none of these can be conclusively identified as being related to the compressed core feature. Differentiating protons by energy (based on the size of the pits they create in the CR-39 detector) did not change this conclusion. It can therefore only be said that these radiographs are consistent with the presence of a compressed axial magnetic field, but are not sufficient to determine the magnitude of the compressed field.

\section{Direct Inversion of Proton Radiographs\label{sec:results_inversion}}

\begin{figure*}
\includegraphics[width = 0.7\textwidth]{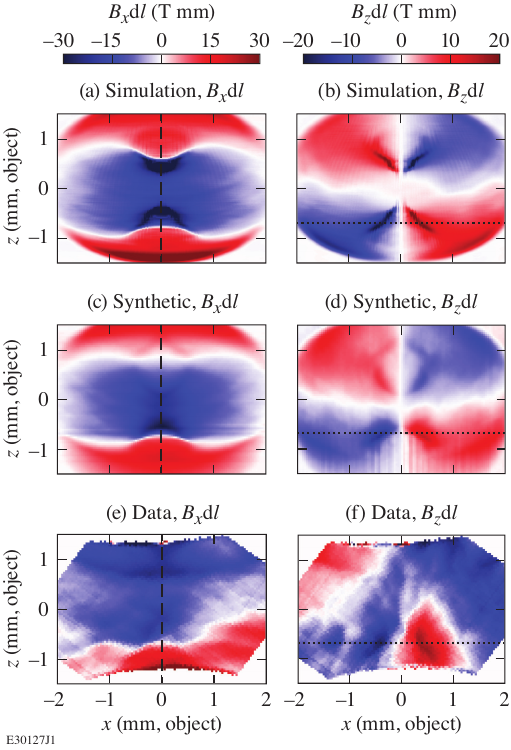}
\caption{\label{inversion_comparison_2D} (a,b) The line-integrated horizontal ($B_x$) and vertical ($B_z$) fields from the unmagnetized simulation of Exp.~1. (c,d) The same quantities retrieved by the power-diagram algorithm from a synthetic radiograph generated with the same simulated fields. (e,f) The same quantities retrieved using the power-diagram algorithm from the unmagnetized experimental radiograph. The inversion results are expressed entirely in terms of a magnetic field; however, some features in the experimental radiograph may actually be created by electric fields that are not included in the synthetic radiographs.
}
\end{figure*}

\begin{figure}
\includegraphics[width = 0.45 \textwidth]{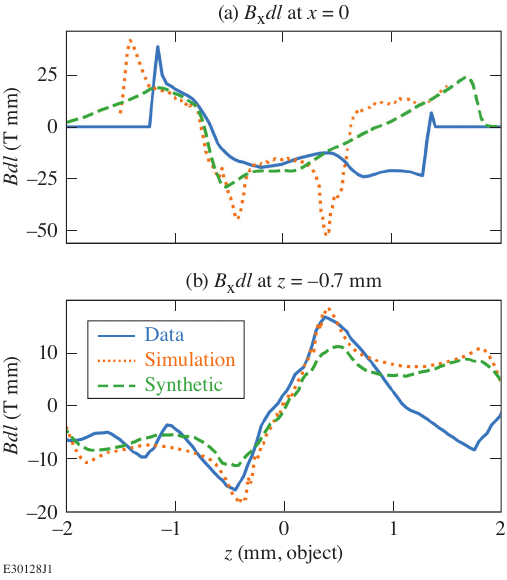}
\caption{\label{inversion_comparison_lineouts} Comparisons of lineouts through the simulated line-integrated fields and reconstructed fields in Fig.~\ref{inversion_comparison_2D}. (a) and (b) correspond to to the vertical dashed lines and the horizontal dotted lines in Fig.~\ref{inversion_comparison_2D} respectively.
}
\end{figure}

In the small deflection limit (where proton trajectories do not cross) it is theoretically possible to invert a proton radiograph to recover the 2-D line-integrated force. Fundamentally, this is an optimal transport problem, finding the pattern of deflections necessary to transform the source profile into the measured radiograph. A number of algorithms have been published and made publicly available to perform these inversions, with approaches ranging from solving a Poisson-like differential equation~\cite{Bott2017proton}, statistical reconstruction~\cite{Kasim2019retrieving}, and computational geometry~\cite{Kasim2017quantitative}. All of these but the statistical reconstruction method require that the normalized proton source profile be known. However, in these experiments this can be well-approximated as uniform because the protons produced by the D$^3$He backlighter sources are close to isotropic. All of the inversion algorithms assume that all of the protons in the original source profile are collected somewhere on the radiograph. 

When deflections are not small (e.g. proton trajectories cross) there is no longer a unique solution to the inversion problem and nonlinear caustic features begin to appear in the radiographs. To evaluate how far these algorithms can be extended into this regime, we benchmarked several algorithms on a suite of test problems~\cite{Davies2022evaluation}. We found that most of the algorithms failed to retrieve the correct line-integrated fields soon after the first caustics appeared anywhere in the image. However, one computational geometry algorithm~\cite{Kasim2017quantitative}, hereafter referred to as the ``power-diagram algorithm", continued to retrieve a reasonable approximation of the correct line-integrated field profile (albeit with a magnitude lower than the true value) in the presence of much larger deflections. This algorithm finds a solution corresponding to the smallest possible deflections necessary to reproduce a given radiograph. Critically, this algorithm is also capable of accurately reconstructing portions of the radiograph with low deflections even while other regions contain large deflections.

Figure~\ref{inversion_comparison_2D} shows the results of applying the power-diagram algorithm to both the unmagnetized experimental and synthetic radiographs shown in Fig.~\ref{radiographs}, alongside the line-integrated magnetic fields from the \textit{HYDRA} simulation for comparison. A more-quantitative comparison of lineouts from each subfigure is shown in Fig.~\ref{inversion_comparison_lineouts}. Inverting the synthetic radiograph has fewer complications than inverting the experimental radiograph since the synthetic radiograph captures all of the deflected protons and the source profile is uniform (since no coil shadows are included). As a result, this inversion accurately recovers the general spatial profile of the line-integrated fields [Figs.~\ref{inversion_comparison_2D}(c) and~\ref{inversion_comparison_2D}(d)]. Some of the sharp features in the line-integrated fields [particularly dips in $B_x \text{d}l$ in Fig.~\ref{inversion_comparison_lineouts}(a)] are missing in the reconstructions, because proton trajectories across these sharp field gradients cross and the inversion algorithm finds the minimum deflection solution assuming that trajectories do not cross.

Inverting the experimental radiographs is more complicated. In these radiographs many protons are deflected off of the detector or blocked by the coils, violating one assumption of the inversion algorithm. As a result, the inversions [Figs.~\ref{inversion_comparison_2D}(e) and~\ref{inversion_comparison_2D}(f)] are less accurate, especially near the edges where more protons are lost. The inversions show strong fields near the surfaces of the coils, which are likely electric fields caused by charging of the coil surfaces. The shadow of the coils also complicates the source profile. To compensate, the source profile is modeled as uniform between the coil shadows and zero within the shadows. Despite these issues, the inversion of the unmagnetized experimental radiograph does reproduce some features reminiscent of the simulated line-integrated fields and accurately matches the magnitude of these fields (Fig.~\ref{inversion_comparison_lineouts}). 

These problems are even more severe in the case of the magnetized experimental radiographs. In this case the coil shadows are larger and the source profile between the coils is also distorted by the field. Consequently, our inversions of the magnetized experimental radiograph did not reproduce the features of the corresponding line-integrated simulated magnetic field.

\section{Conclusions\label{sec:conclusion}}

Two experiments on the OMEGA Laser System used proton radiography to measure magnetic fields in magnetized cylindrical implosions based on the miniMagLIF platform~\cite{Davies2017laser}. Although the goal of the experiments was to measure the magnitude of the compressed axial magnetic field in the core of the implosion, this field was obfuscated by two features in the coronal plasma produced by the compression beams: an azimuthal self-generated magnetic field and small-scale, high-amplitude structures attributed to collisionless effects in the coronal plasma. 

To understand these features, we developed an open-source particle-tracing algorithm to generate synthetic radiographs from simulated fields produced by 3-D \textit{HYDRA} simulations~\cite{PlasmaPyv07}. These synthetic radiographs reproduce many of the features of the experimental radiographs with the exception of small-scale structures, which we attribute to kinetic or multi-species effects that are not present in the MHD simulations. A direct inversion algorithm successfully inverts a synthetic radiograph but retrieves limited information from the experimental radiographs, in part because not all protons are collected on the detector. We conclude that these experimental radiographs are consistent with the presence of a compressed axial field, but that a measurement of the compressed field is prevented by the self-generated azimuthal magnetic fields in the coronal plasma and the small-scale structure fields. 

\edit{}{MHD simulations predict that conditions in the coronal plasma are non-local (Sec.~\ref{sec:simulations}), and previous authors have described how non-local effects can suppress the Bierman mechanism and reduce the magnitude of self-generated magnetic fields~\cite{Sherlock2020suppression, Ridgers2021inadequacy, Campbell2022measuring}. In the regime predicted by the simulations shown in Fig.~\ref{hydra} ($\lambda_{ei} \sim 0.01 \text{ to } 1$), the model published by Sherlock and Bissell~\cite{Sherlock2020suppression} (which is developed for a similar plasma parameter regime) predicts that the Biermann field growth rate will be reduced by a factor of between 2 to 10. However, the lower end of this prediction falls within the significant uncertainty of our measurement of the magnitude of the magnetic fields in these experiments (see Sec.~\ref{sec:results_quantitative}), so we are unable to test this prediction. We note, however, that the position of the radiograph features (which are known to higher certainty than the magnitude of the fields) agrees well with the synthetic radiographs from simulations, suggesting that at least the location and shape of the fields are well simulated by MHD.}

A previous experiment on the Omega Laser System (Gotchev \textit{et al.}) ~\cite{Gotchev2009laser, Knauer2010compressing} successfully made measurements of the compressed axial magnetic field in a similar cylindrical implosion without similar issues with self-generated azimuthal or small-scale structure fields. Comparing the design of this previous experiment to the current work provides guidance for the design of future experiments. The diameter of the cylindrical target in Gotchev \textit{et al.} was larger than in the current work while the laser pulse length was shorter (Table~\ref{setup_table}). As a result, the time between maximum convergence (when the radiographs were recorded) and the end of the drive pulse was longer, allowing the coronal plasma and self-generated azimuthal fields to dissipate before the proton radiograph was taken. Additionally, the proton radiography performed in Gotchev \textit{et al.} was normally incident ($\theta_\text{T}=0$), minimizing the effect of any azimuthal fields that were present and allowing the radiograph to be integrated along the cylinder axis to improve the signal-to-noise ratio. The coils used in the Gotchev \textit{et al.} experiment were also larger and further from the cylinder. This may have reduced the significance of charging of the coil surfaces or ablation from those surfaces, and may explain why Gotchev \textit{et al.} did not observe small-scale structure. 

In many experiments, the ability to change radiography angle $\theta_\text{T}$, target dimensions, laser pulse duration, and coil geometry is limited by other design considerations. However, future attempts to measure compressed axial magnetic fields in cylindrical implosions should include among these considerations the potential impact of self-generated fields on the measurement. Future proton radiography experiments in which a significant number of protons are blocked or deflected off of the detector should also include a separate measurement of the source flux ($I_0$) on an unobstructed line of sight. Finally, future experiments should consider the potential of proton radiography at oblique incidence to make axially resolved measurements of azimuthal magnetic fields.

\begin{acknowledgments} 
This material is based upon work supported by the Advanced Research Projects Agency-Energy (ARPA-E) under Award Number DE-AR0000568, the Department of Energy National Nuclear Security Administration under Award Numbers DE-NA0003856 and DE-SC0020431, the University of Rochester, and the New York State Energy Research and Development Authority. The data that support the findings of this study are available from the corresponding author upon reasonable request.

This report was prepared as an account of work sponsored by an agency of the U.S. Government. Neither the U.S. Government nor any agency thereof, nor any of their employees, makes any warranty, express or implied, or assumes any legal liability or responsibility for the accuracy, completeness, or usefulness of any information, apparatus, product, or process disclosed, or represents that its use would not infringe privately owned rights. Reference herein to any specific commercial product, process, or service by trade name, trademark, manufacturer, or otherwise does not necessarily constitute or imply its endorsement, recommendation, or favoring by the U.S. Government or any agency thereof. The views and opinions of authors expressed herein do not necessarily state or reflect those of the U.S. Government or any agency thereof.
\end{acknowledgments}

\end{document}